\documentclass[aps,prl,twocolumn,showpacs]{revtex4}
\usepackage{amssymb}
\usepackage{amsmath}
\usepackage{color}
\usepackage{lipsum}
\usepackage{titlesec}
\usepackage{wrapfig}
\usepackage{color}
\usepackage[export]{adjustbox}
\usepackage{graphicx}
\usepackage{epstopdf}
\newcommand{\sech}{\rm sech}

\titleformat{\section}
   {\bfseries}{\thesection}{}{}
   
\titleformat{\subsection}
   {\bfseries\small}{\thesection}{}{}

\begin{document}
\title{All-phononic Amplification in Coupled Cantilever Arrays based on Gap Soliton Dynamics}
\author {Merab Malishava}
\affiliation {Department of Physics, I. Javakhishvili Tbilisi State
University, 3 Chavchavadze, 0179 Tbilisi, Georgia}

\begin{abstract}
	We present a mechanism of amplification of phonons by phonons on the basis of nonlinear band-gap transmission phenomenon. As a concept the idea may be applied to the various number of systems, however we introduce the specific idea of creating amplification scenario in the chain of coupled cantilever arrays. One chain is driven at the constant frequency located in the upper band of the ladder system, thus no wave enters the system. However the frequency is specifically chosen to be very close to the maximum value of frequency corresponding to dispersion relation of the system. Amplification scenario happens when a counter phase pulse of same frequency with a small  amplitude is introduced to the second chain. If both signals exceed a threshold amplitude for the band-gap transmission a large amplitude soliton enters the system - therefore we have an amplifier. Although the concept may be applied in a variety of contexts - all optical or all-magnonic systems, we choose the system of coupled cantilever arrays and represent a clear example of the application of presented conceptual idea. Logical operations is the other probable field, where such mechanism could be used, which might significantly broaden the horizon of considered applications of band-gap soliton dynamics.
\end{abstract}

\pacs{05.45.-a, 43.25.+y, 05.45.Yv} \maketitle

\section{Introduction}

The first documented observations of soliton waves occurred in 1834 by John Scott Russell, although the significance of soliton waves became clear later, with the studies of Korteweg - de Vries equation, ultimately brought mathematical clarity to the processes observed before.

As the studies on nonlinear phenomena went on, the numerical experiments on discrete nonlinear structures emerged. The first of those is known to be conducted by Fermi, Pasta and Ulam in 1954 \cite{fpu}. The studies on FPU model and its developments \cite{klein,braun,toda} led to the discovery of solitons \cite{flach0,zabusky,thierry1}. The model of anharmonic oscillator chains became a strong tool for modeling and explaining phenomenas in various branches of physics and contributed to the fundamentals nonlinear wave phenomena \cite{scott,flach01} as well as statistical physics \cite{izrailev,ruffo1}, has been applied to explain thermal conductivity in various physical systems \cite{kaburaki, bambi}, contributed to understanding the interrelation between integrability and chaos \cite{chaos, chaos1}, was used as a model for representing complex condensed matter systems \cite{flach1,flach2} and electric transmission lines \cite{trans,trans1}. 
 
 The studies of phononics and advancements in phonon laser technology led to researches on phonon diodes  \cite{laser,laser1, diode, diode1,diode2,diode3} and all-phonon transistors \cite{ all, acoust,merab}.

In this  article we are going to consider a system of coupled cantilever arrays and apply non-linear band gap transmission\cite{leon, ramaz1, ramaz2} in order to achieve the amplification of weak acoustic waves.

\begin{figure}[h]
\includegraphics[width=0.5\textwidth, left]{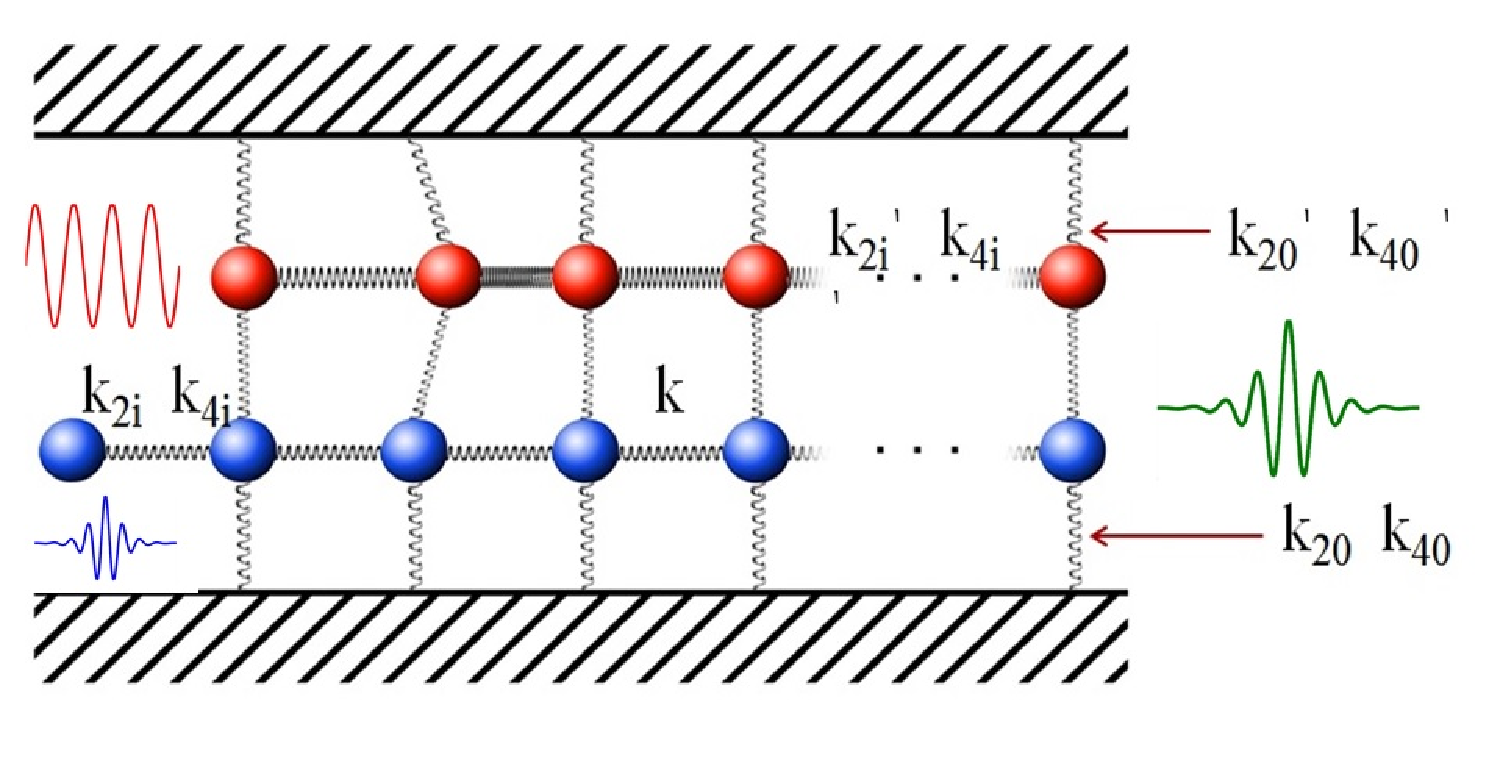}
\caption{We have coupled FPU chains to represent the system of coupled cantilever arrays - the units of the chains represent the cantilevers, while the means of linking cantilevers one to another are represented by strings. Its also worth mentioning that we consider the system where any particular unit is not linked solely to its neighbors, but to any number of units in its neighborhood. The upper chain - red is driven with a constant frequency below the gap transmission value. A signal with a small amplitude is introduced to the bottom chain - blue, chosen so that the amplitude of both signals is enough to exceed the threshold value and a large amplitude soliton enters the system - green curve.}
\label{model1}

\end{figure}
\begin{figure}[h]
\includegraphics[width=0.5\textwidth, left]{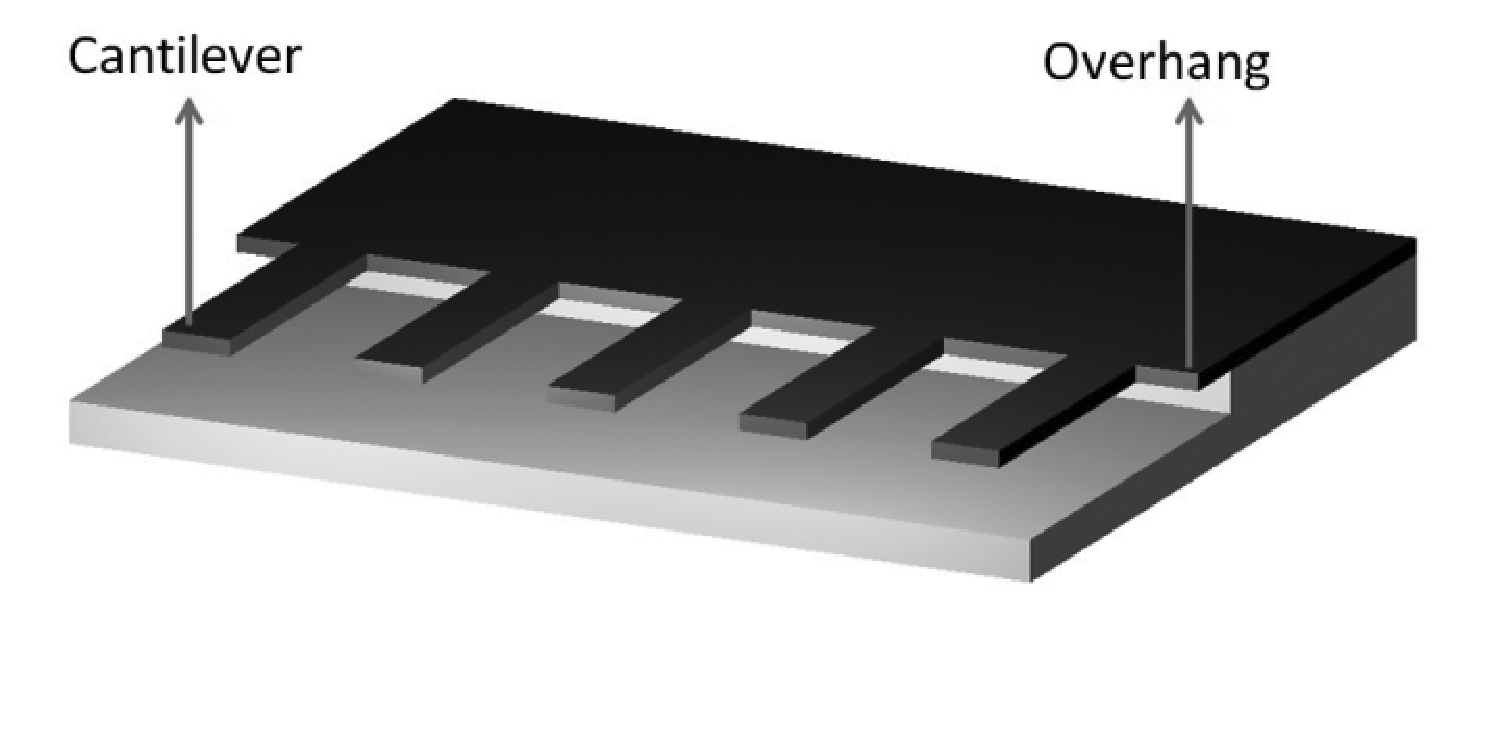}
\caption{The scheme of a mono-element cantilever array. Exactly same cantilevers are linked by means of the overhang, thus any particular cantilever can be represented as an oscillator.}
\label{Cantilever1}
\end{figure}

We study mono-element cantilever arrays, which consist of same cantilevers, which are connected to neighbors by means of the overhang, so that any particular cantilever can be observed as an oscillator [see Fig.\ref{Cantilever1}]. A number of works on wave propagation as well as logic operation tools have been made on the basis of cantilever arrays system \cite{canti, canti1, canti2}.
We are going to represent the system of coupled cantilever arrays by introducing a model of coupled FPU chains \cite{sievers1}, with on-site terms [Fig.\ref{model1}]; we also consider a system where any particular oscillator is linked to any number of oscillators in its neighborhood, although we are going to use six neighbors in numerical experiments.  The idea behind the mechanism is driving one upper chain with a constant frequency just below the band gap transmission, while the bottom chain is at rest. We then introduce a pulse with a small amplitude to the lower chain with phase specifically chosen so that the overall amplitude of both signals is enough to exceed the threshold value. As a result large amplitude soliton enters into the system, thus we have amplification of a small acoustic signal.  

\section{Deriving Analytical Solution for the Problem}
\subsection{Introducing Equations of Motion}
We begin with the Hamiltonian for system of coupled ladders with $N$ units each:
\begin{eqnarray}
H=H_{u}+H_{w}+\frac{k}{2}(w_n-u_n)^2
\end{eqnarray}
Where $H_{u}$:
\begin{eqnarray}
H_{u} = \sum\limits_{n=1}^N\bigg[\frac{m{\dot u_n}^2}{2}+
\sum\limits_{i=-N_1}^{N_1} \frac{k_{2i}}{2}(u_{n+i}-u_n)^2+
\nonumber \\
\sum\limits_{i=-N_1}^{N_1}\frac{k_{4i}}{4}(u_{n+i}-u_n)^4+
\frac{k_{20}}{2}u_n^2+\frac{k_{40}}{4}u_n^4\bigg]
\end{eqnarray}
where $m$, $k$, $k_{2i}$, $k_{4i}$, $k_{20}$, $k_{40}$ are the
parameters of the chain, namely, masses of units and stiffness
coefficients of springs. Note that $H_{w}$ has the identical form
with just other parameters except of $m$, which we consider the same
for units in both ladders.We should point out that $N_1$
representing the number of units which are considered to interact
with unit $u_n$ may vary. This fact brings up the possibility of
describing the whole variety of systems using the pattern which is
going to be considered below.

The equations of motion for $n$-th unit in each chain corresponding
to the Hamiltonian $H$ will have the form of:
\begin{eqnarray}
m\ddot u_n &=&\sum\limits_{i=-N_1}^{N_1}k_{2i}(u_{n+i}-u_n)+\sum\limits_{i=-N_1}^{N_1}k_{4i}(u_{n+i}-u_n)^3 \nonumber \\
&&
+k_{20}u_n+k_{40}u_n^3+k(w_n-u_n) \nonumber \\
m\ddot w_n &=&k_1^\prime\sum\limits_{i=-N_1}^{N_1}k_{2i}^\prime(w_{n+i}-w_n)+\sum\limits_{i=-N_1}^{N_1}k_{4i}^\prime(w_{n+i}-w_n)^3 \nonumber \\
&& +k_{20}^\prime w_n+k_{40}^\prime w_n^3+k(u_n-w_n)\label{1}
\end{eqnarray}

\subsection{Deriving the Solution}
We use well-known approach, seeking the solution in a form of the
following perturbative expansion:

\begin{eqnarray}
\textbf{U}=\sum_{\alpha=1}^\infty\epsilon^\alpha
\sum_{m=-\infty}^{+\infty}{\textbf{U}_m^{(\alpha)}(\tau,\xi)e^{im(pn-\Omega
t)}} \label{exp}
\end{eqnarray}
where we define column vector
$\textbf{U}^{(\alpha)}=(u_n^{(\alpha)}, w_n^{(\alpha)})$, while
$\xi$ and $\tau$ are slow variables introduced through:
$\xi=\epsilon(n-vt)$ and $\tau=\epsilon t^2$; $v$ is a soliton group
velocity defined below and $\epsilon$ is a small expansion
parameter.

We go on with equating powers of $\epsilon$ substituting expansion
(\ref{exp}) in set of equations (\ref{1}). In the linear
approximation we have the column vector $\textbf{U}_1^{(1)}\equiv
\left( u_n^{(1)},w_n^{(1)}\right)=\varphi(\xi,\tau)\mathbf{R}$ and
$\textbf{U}_m^{(1)}=0$ for $|m|\neq1$; not restricting generality we
can take a space-time independent column vector as
$\mathbf{R}=\left(R,1\right)$, where $R$ is a complex number and
$\varphi(\xi,\tau)$ is a scalar function of slow variables to be
determined in the next approximations. Then by considering
$\alpha=1$ (linear approximation) and the harmonic $m=1$ we arrive
to the equation:
\begin{eqnarray}
\hat{\mathbf{W}}*\mathbf{R}=0 \label{WW}
\end{eqnarray}
where
\begin{eqnarray}\hat{\mathbf{W}}=\left(
\begin{array}{cc}
m\Omega^{2}+2s_p-k_{20}-k ~~~~~~~~~~~~~~ k~~~~~~~~~~~~~ \\
~~~~~~~~~~~k ~~~~~~~~~~~~~~ m\Omega^{2}+2l_p-k_{20}^\prime-k
\end{array}
\right) \nonumber
\end{eqnarray}
with
\begin{eqnarray}
s_p=\sum\limits_{s=1}^{N_1}k_{2s}(\cos sp-1),
l_p=\sum\limits_{l=1}^{N_1}k_{2l}^\prime(\cos lp-1)\nonumber
\end{eqnarray}
The solvability of this equation demands Det($\hat{\mathbf{W}})=0$,
which gives us two branches of dispersion relations:
\begin{eqnarray}
\Omega^2_{1,2}&=&\frac{1}{m}\bigg[-s_p-l_p+\frac{1}{2}(k_{20}+k_{20}^\prime)+k\pm \nonumber \\
&\pm&\sqrt{(s_p-l_p+\frac{1}{2}(k_{20}-k_{20}^\prime))^2+k^2}\bigg]
\label{disp}
\end{eqnarray}

\begin{figure}[h]
\includegraphics[width=0.5\textwidth, left]{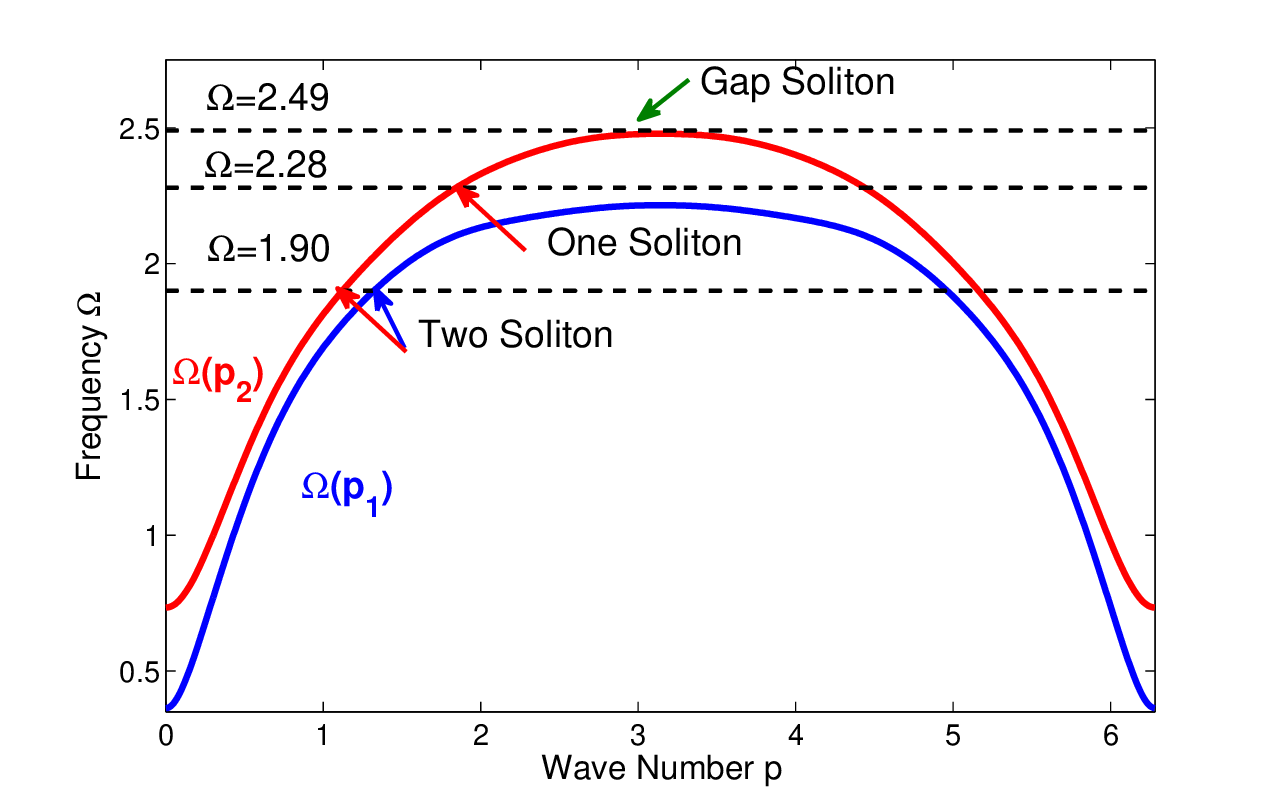}
\caption{The dispersion relation for the considered system of two weakly coupled cantilever arrays. The blue and red curves represent dispersion relations for $p_1$ and $p_2$ respectively. As a result of these relations we consider three regimes: a)Two Soliton (lower dashed line), b)One soliton (middle dashed line), c) Amplification scenario (upper dashed line), with respective frequencies.}
\label{dispersia}
\end{figure}

As a result of \eqref{disp} we obtain two corresponding column vectors $\mathbf{R}_j=(R_j,1)$ with
$R_j$ expressed with the linear parameters of the problem
$R_j=k/[k+k_{20}-m\Omega^2_j-2s_p]$, where $j=1,2$. Next we
introduce a row vector $\mathbf{L}=(L,1)$ through the equation
$\mathbf{L}*\hat{\mathbf{W}}=0$, that gives us two row vectors
$\mathbf{L}_j$. In our case the respective components of row
$\mathbf{L}_j$ and column $\mathbf{R}_j$ are identical
$L_j=R_j$.Thus in linear limit we have following matrix relations:
\begin{eqnarray}
\hat{\mathbf{W}}\left(\Omega_j\right)*\mathbf{R}_j=0, \qquad
\mathbf{L}_j*\hat{\mathbf{W}}\left(\Omega_j\right)=0. \label{WW1}
\end{eqnarray}
In the following for presentation clarity we omit the indexes $j$
and restore them at the end of the calculations. We go on with a
second approximation ($\alpha=2$) substituting again \eqref{exp}
into \eqref{1} and considering first harmonic $m=1$, which leads us
to the following equation:
\begin{eqnarray}
\hat{\mathbf{W}}\textbf{U}^{(2)}+2i(\hat{\mathbf{B}}-m\Omega v
\hat{\mathbf{I}})\frac{\partial \varphi}{\partial \xi}\mathbf{R}=0,
\label{222}
\end{eqnarray}
where
\begin{eqnarray} \hat{\mathbf{B}}=\left(
\begin{array}{cc}
\sum\limits_{s=1}^{N_1}sk_{2s}\sin sp& 0 \\ 0 &
\sum\limits_{l=1}^{N_1}lk_{2l}^\prime\sin lp
\end{array}
\right)
\end{eqnarray}
Then multiplying \eqref{222} by $\mathbf{L}$ one has
\begin{eqnarray}
\mathbf{L}(\hat{\mathbf{B}}-m\Omega v
\hat{\mathbf{I}})\frac{\partial \varphi}{\partial \xi}\mathbf{R}=0.
\label{A3}
\end{eqnarray}
In order to identify constant $v$ in the equation above, let us take
the derivative of \eqref{WW} over $p$ and multiply then on the row
vector $\mathbf{L}$. One gets:
\begin{eqnarray}
\mathbf{L}\frac{\partial \hat{\mathbf{W}}}{\partial
p}\mathbf{R}=2\mathbf{L}\left(m\frac{d\Omega}{dp}\Omega\hat{\mathbf{I}}-
\hat{\mathbf{B}}\right)\mathbf{R}=0. \label{A4}
\end{eqnarray}
Comparing now \eqref{A3} and \eqref{A4} we immediately get the
equality $v=\partial \Omega/\partial p$, thus the definition for
group velocity, while from \eqref{222} one can solve
$\textbf{U}^{(2)}$ as follows:
\begin{eqnarray}
\textbf{U}^{(2)}=-2i\hat{\mathbf{W}}^{-1}(\hat{\mathbf{B}}-m\Omega v
\hat{\mathbf{I}})\frac{\partial \varphi}{\partial \xi}\mathbf{R}.
\label{A5}
\end{eqnarray}
In the third approximation, equating powers of $\epsilon$ for
$\alpha=3$ and first harmonic $m=1$ we have:
\begin{eqnarray}
\hat{\mathbf{W}}\textbf{U}^{(3)}+2i(\hat{\mathbf{B}}-m\Omega v
\hat{\mathbf{I}})\frac{\partial \textbf{U}^{(2)}}{\partial
\xi}+2im\Omega \frac{\partial\varphi}{\partial \tau}{\mathbf{R}}- \\
\nonumber \label{A11}
-\hat{\mathbf{C}}\frac{\partial^2\textbf{U}^{(1)}}{\partial\xi^2}+12\hat{\mathbf{P}}{\mathbf{N}}|\varphi|^2\varphi=0
\end{eqnarray}
where
\begin{eqnarray}
\hat{\mathbf{C}} = mv^2\hat{\mathbf{I}}- \left(
\begin{array}{cc}
\sum\limits_{s=1}^{N_1}s^2k_{2s}\cos sp ~~~~~0 ~~~~~~~\\
~~~0~~~~ \sum\limits_{l=1}^{N_1}l^2k_{2l}^\prime\cos lp
\end{array}\right) \nonumber
\end{eqnarray}

\begin{eqnarray}
\hat{\mathbf{P}} = \left(
\begin{array}{cc}
\sum\limits_{s=1}^{N_1}k_{4s}(1-\cos sp)^2 + k_{40} ~~~~~0 ~~~~~~~\\
~~~0~~~~ \sum\limits_{l=1}^{N_1}k_{4l}^\prime(1-\cos lp)^2 +
k_{40}^\prime
\end{array}\right) \nonumber
\end{eqnarray}

\begin{eqnarray}
{\mathbf{N}} = \left(\begin{array}{cc} R^3 \\ 1
\end{array} \nonumber
\right)
\end{eqnarray}
Now noting that
\begin{eqnarray}
2\left(\hat{\mathbf{B}}-m\Omega v
\hat{\mathbf{I}}\right)\equiv-\frac{\partial
\hat{\mathbf{W}}}{\partial p}; \nonumber \\ \quad
\hat{\mathbf{C}}\equiv\frac{1}{2}\frac{\partial^2
\hat{\mathbf{W}}}{\partial p^2}-m\Omega\frac{\partial^2
\Omega}{\partial p^2}\hat{\mathbf{I}} \label{A91}
\end{eqnarray}
We can further simplify \eqref{A11} multiplying it on
$\mathbf{L}$ and taking into account \eqref{A5} and \eqref{A91}:
\begin{eqnarray}
\mathbf{L}\left(m\Omega\frac{\partial^2 \Omega}{\partial
p^2}\hat{\mathbf{I}}+\frac{\partial \hat{\mathbf{W}}}{\partial
p}\hat{\mathbf{W}}^{-1}\frac{\partial \hat{\mathbf{W}}}{\partial
p}-\frac{1}{2}\frac{\partial^2 \hat{\mathbf{W}}}{\partial
p^2}\right)\textbf{R}\frac{\partial^2\varphi}{\partial \xi^2}
\nonumber \\
+2im\Omega \frac{\partial\varphi}{\partial \tau}\mathbf{L}\mathbf{R}
+12\mathbf{L}\hat{\mathbf{P}}{\mathbf{N}}|\varphi|^2\varphi=0
\label{A61}
\end{eqnarray}
We can get a final form for \eqref{A61} taking first and second
derivatives of Eq. \eqref{WW} over $p$:
\begin{eqnarray}
\frac{\partial \hat{\mathbf{W}}}{\partial
p}\mathbf{R}+\hat{\mathbf{W}}\frac{\partial \textbf{R}}{\partial
p}=0; \label{A8} \\ \frac{\partial^2 \hat{\mathbf{W}}}{\partial
p^2}\mathbf{R}+2\frac{\partial \hat{\mathbf{W}}}{\partial
p}\frac{\partial \textbf{R}}{\partial
p}+\hat{\mathbf{W}}\frac{\partial^2 \textbf{R}}{\partial p^2}=0
\nonumber
\end{eqnarray}
Solving now $\partial \textbf{R}/\partial p$ from the first equation
and substituting it in the second one and then multiplying it on
$\textbf{L}$ one gets the following relation:
\begin{eqnarray}
\mathbf{L}\frac{\partial \hat{\mathbf{W}}}{\partial
p}\hat{\mathbf{W}}^{-1}\frac{\partial \hat{\mathbf{W}}}{\partial
p}\textbf{R}-\frac{1}{2}\mathbf{L}\frac{\partial^2
\hat{\mathbf{W}}}{\partial p^2}\mathbf{R}=0, \label{A9}
\end{eqnarray}
and now substituting this into the \eqref{A61} and restoring indexes
$j$-s one finally arrives to the Nonlinear Schr\"odinger (NLS)
Equation for two nonlinear modes $j=1,2$:
\begin{eqnarray}
2i\frac{\partial\varphi_j}{\partial \tau}+\Omega_j^{\prime
\prime}\frac{\partial^2\varphi_j}{\partial\xi^2}-\Delta_j|\varphi_j|^2\varphi_j=0
\label{nls}
\end{eqnarray}
where
\begin{eqnarray}
\quad\Omega_j^{\prime\prime}={\frac{\partial^2\Omega_j}{\partial
p^2}}\biggr|_{p=p_j}, \nonumber \\
\Delta_j=\frac{1}{m\Omega_j(1+R_j^2)}\bigg[12R_j^4\sum\limits_{s=1}^{N_1}k_{4s}(1-\cos sp)^2 + \nonumber \\
 \nonumber \\ 
+ 12\sum\limits_{l=1}^{N_1}k_{4l}^\prime(1-\cos lp)^2+R_j^4k_{40}+k_{40}^\prime\bigg] \label{nls0}
\end{eqnarray}
and wavenumbers $p_j$ are the solutions of respective dispersion
relations:

\begin{eqnarray}
\Omega=\Omega_j+\Delta_j A_j^2/4
\label{Omega_th}
\end{eqnarray}

The solutions of \eqref{A9} with respect to modes have solitonic
form \cite{taniuti,oikawa,ramaz4}
\begin{eqnarray}
\left(u_n^j, w_n^j\right)=\left(R_j,~~1\right)\frac{A_j\cos(\Omega
t-p_jn)}{\cosh\left[(n-v_jt)/\Lambda_j\right]} \label{7}
\end{eqnarray}
where $A_j$ is a soliton amplitude, while soliton width $\Lambda_j$:
\begin{eqnarray}
\Lambda_j=\frac{1}{A_j}\sqrt{\frac{2\Omega_j^{\prime\prime}}{\Delta_j}}
\end{eqnarray}

\section{Numerical Experiments}
\subsection{Parameters}

For the purpose of numerical experiments we are going to consider dimensionless parameters. We divide \eqref{1} by $k_{21}$ and introduce the following transformations:

\begin{eqnarray}
x \longrightarrow x\sqrt{k_{21}/k_{41}}, t \longrightarrow \sqrt{m/k_{21}}
\label{trans}
\end{eqnarray}

After that we rescale the parameters of the chain and consider new dimensionless $k_{21}=1$ and $k_{41}=1$. Using \eqref{trans} we obtain a new set of parameters (Table \ref{parameters}). Note the real parameters of the chain: $k_{21}=0.0828 kg/s^2$, $k_{41}=4.0\times10^{10} kg/s^2m^2$  $m=7.67\times10^{-13}kg$. Thus by considering \eqref{trans} and these parameters one can obtain the actual characteristics of the chain.
 
\begin{table}[h]
\centering	
\caption{We represent the parameters of considered coupled FPU chains: mass - m, $k_{20}$, $k_{40}$ - linear and qubic onsite terms respectively, $k_{21-26}$, $k_{41-46}$,  linear and qubic intersite terms, as we consider six nearest neighbor oscillators of the every unit of the chain, $k$ - interchain term. Note that all of the parameters are obtained based on the characteristics of cantilever arrays \cite{sievers1}}																		
\label{parameters}
\begin{tabular}{l|l|lll}
\hline
 Parameter &  Chain no.1 &  Chain no.2  &  \\ 
 \hline
 $m$  & 1  & 1 &  \\
 &  &  &  &		\\
 $k_{20}$  & 0.1 & 0.17 &  \\
 &  &  &  &		\\
 $k_{21-26}$  & 1, 0.3720 &  1.3, 0.3297 &  \\
 & 0.1304, 0.0489 & 0.1075, 0.0562 &  &			\\
 & 0.0300, 0.0100 & 0.0272, 0.0106 &  &			\\
 &  &  &  &		\\
 $k_{40}$  & 0.2 & 0.7 &  \\
 &  &  &  &		\\
 $k_{41-46}$  & 1.0000, 0.3725 & 3.500, 1.3900 &  \\
 & 0.1305, 0.0488 & 0.5771, 0.1505 &  &			\\
 & 0.0300, 0.0100 & 0.0807, 0.0395 &  &  \\
 &  &  &  &		\\		
  $k$ & 0.2 & 0.2 &\\
  \hline
\end{tabular}
\end{table}

\subsection{Combining Solutions}
\begin{figure}[h]
\includegraphics[width=0.5\textwidth, left]{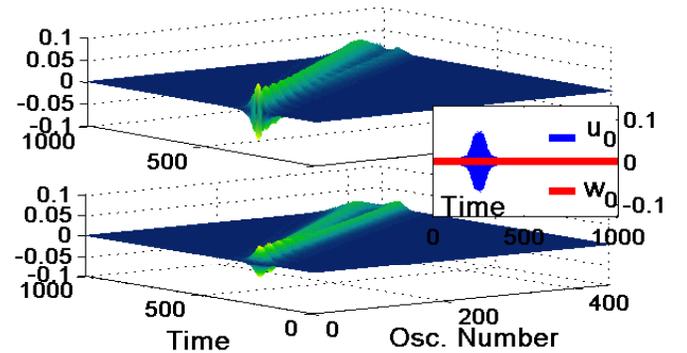}
\caption{The two soliton regime: Combining solutions corresponding to two different modes(lower dashed line in [Fig.\ref{dispersia}]), we generate a soliton wave through the upper chain, while the lower chain is being at rest. As a result due to different velocities solitons depart from each other and propagate independently through the system. Corresponding frequency and amplitude: $\Omega=1.90, A=0.025$}
\label{2Soli}
\end{figure}
Strictly speaking the linear combination of the solutions \eqref{7}
of different $j=1$ and $j=2$ modes is not a solution of the initial
nonlinear problem \eqref{1}, however, in weakly nonlinear limit
(small soliton amplitudes $A_j\ll 1$) and large relative group
velocities $\left|v_1-v_2\right|/v_{1,2}\gtrsim 1$ one can combine
the solutions \eqref{7} acquiring additional phase shift
\cite{oikawa} which could be safely neglected in the mentioned
limits. By this one is able to construct the solution, which
describes the initial excitation of the boundary of the solely upper
chain. In particular, if one takes $A_1=A_2$ and finds such an
excitation frequency that $v_1/\Lambda_1=v_2/\Lambda_2$, the
combination $\left(u_n^1, w_n^1\right)-\left(u_n^2, w_n^2\right)$ at
the origin $n=0$ gives
\begin{eqnarray}
\left(u_0^1, w_0^1\right)-\left(u_0^2,
w_0^2\right)=\left(R_1-R_2,~~0\right)\frac{A_1\cos(\Omega
t)}{\cosh\left[v_1t/\Lambda_1\right]}, \label{9}
\end{eqnarray}
thus driving both chains in time according to the above expression
one can excite two soliton solution belonging to different branches.
That is displayed in [Fig.\ref{2Soli}], driving in numerical simulations the
left end of the upper chain $u_0$ with a frequency $\Omega=1.90$ and
amplitude $A=\left(R_1-R_2\right)A_1$ with $A_1=0.025$ and
calculating $R_j$ from Eq. \eqref{222}. At the same time the lower
chain is kept pinned at the left boundary ($w_0=0$) according again
to the expression \eqref{9}. As seen, the numerical test is just in
tact with the expectation, as  far as according to \eqref{7} we
observe different amplitudes for the solitons in the upper chain and
just the same $A_1=0.025$ in the lower one.

\begin{figure}[h]
\includegraphics[width=0.5\textwidth, left]{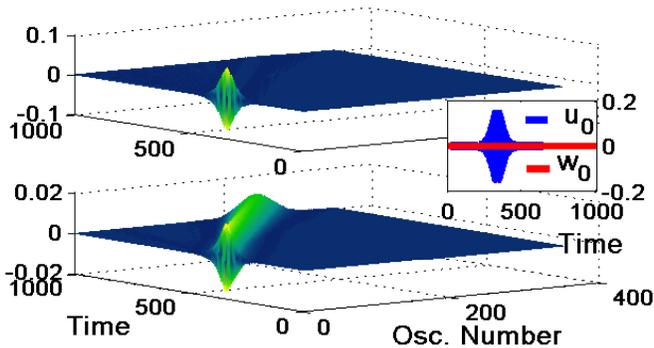}
\caption{The one soliton regime: Combining soliton solution corresponding to $p_1$ with a soliton-like pulse with imaginary $p_2$ \eqref{10} (middle dashed line in [Fig.\ref{dispersia}]). As a result we generate one soliton wave through the upper chain. It then propagates freely into the system, while the first unit of the lower chain $w_0$ is at rest. Corresponding frequency and amplitude: $\Omega=2.28, A=0.025$}
\label{1Soli_meh}
\end{figure}
Next we examine one soliton generation driving again only upper
chain with a frequency lying in the limits
$\Omega_2(\pi)<\Omega<\Omega_1(\pi)$, particularly we apply
$\Omega=2.28$ in numerical simulations (see middle horizontal line
in [Fig.\ref{dispersia}]. In this case antisymmetric mode ($j=1$) solution could
be again presented in solitonic form \eqref{7}, while the symmetric
mode ($j=2$) has no longer a solitonic profile, instead it is
described by evanescent wave since the corresponding wavenumber
$p_2$ is imaginary number (solution of dispersion relation
$\Omega=\Omega_2(p)$ has no real roots):
\begin{eqnarray}
\left(u_n^2,
w_n^2\right)=\left(R_2,~~1\right)B(t)e^{-\left|p_2\right|n}\cos(\Omega_2
t) \label{10}
\end{eqnarray}
where $B(t)$ can slowly vary in time. This means that we observe
only one soliton entering the chain. As we try to nullify
oscillations in the lower chain $B(t)$ should take a form of
$B(t)=A_1\sech\left(\emph{v}_1t/\Lambda_1\right)$ and then the
combination $\left(u_n^1, w_n^1\right)-\left(u_n^2, w_n^2\right)$ at
the origin $n=0$ gives the same form of the driving as in the
previous case \eqref{9} of the two soliton generation. The results
are displayed in [Fig.\ref{1Soli_meh}], and as seen driving the upper chain with a
frequency $\Omega=2.28$ now one monitors the generation of a single
envelope soliton.
\subsection{Numerical Experiments for Amplification Scenario}

\begin{figure}[h]
  \centering
  \begin{minipage}[b]{0.5\textwidth}
    \includegraphics[width=\textwidth]{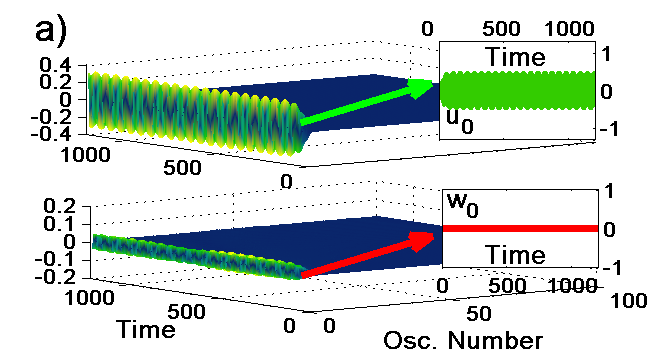}
  \end{minipage}
  \hfill
  \begin{minipage}[b]{0.5\textwidth}
    \includegraphics[width=\textwidth]{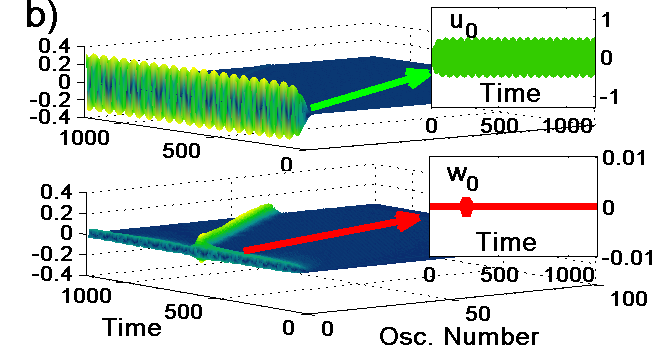}
    \caption{{The amplification scenario: we consider driving frequency of the upper chain at $\Omega=2.49$ with $A=0.5098$, which is slightly above possible frequencies introduced through dispersion relation (upper dashed line in [Fig.\ref{dispersia}]). In both instances the insets show the motion of first unit of the chain - the source.
a) - no wave enters the system as the first unit of the upper chain oscillates with given frequency and amplitude, while there is no excitation introduced to the lower chain. 
b) - a pulse with small amplitude - $A^\prime=0.004$ \eqref{11} is introduced through the lower chain. As a result the band gap soliton enters the system with amplitude of $A_f=0.3274$, thus giving the amplification of $80$ times at the output of the system.}}
    \label{NoSoli}
  \end{minipage}
\end{figure}

Finally we consider the case $\Omega=2.49$ (upper dashed line in
[Fig.\ref{dispersia}]) lying in the band gap of both modes, for which only
evanescent wave solutions \eqref{10} is realized for the modes if
the driving amplitude is small. However, if the amplitude exceeds
some threshold value, a gap soliton can be created and propagate
along the ladder. For the estimation of this threshold value, we
assume that the upper chain is driven with the amplitude $A$ while
the lower one is kept pinned.Then, looking at the typical solution
of such a scenario \eqref{9} one can notice that the weight of the
antisymmetric mode $A_1$ is defined from the relation
$A=A_1\left(1-R_1/R_2\right)$ and the threshold value is calculated
from the expression of nonlinear frequency shift \eqref{Omega_th}. Thus a
threshold amplitude for which driving of the upper chain produces a
gap soliton could be straightforwardly derived as follows:
\begin{eqnarray}
A_{th}=\left(1-R_1/R_2\right)\sqrt{4\left[\Omega-\Omega_1(\pi)\right]/\Delta_1}.
\label{11}
\end{eqnarray}
Determining $A_{th}$ gives us an opportunity to realize the
amplification scenario. For this we create the continuous driving in
the upper chain with a band-gap frequency $\Omega=2.49$ and
amplitude just below the threshold, then even small counter-phase
pulse in the lower chain can help to overcome the threshold and
provide the necessary amplification effect for the weak pulse. 
Thus we choose the driving amplitude and estimate a pulse needed for a single gap soliton to enter the system using \eqref{Omega_th}:
\begin{eqnarray}
A^\prime=R_1\sqrt{4\left[\Omega-\Omega_1(\pi)\right]/\Delta_1}.
\label{11}
\end{eqnarray}
For the numerical experiment displayed in the [Fig.\ref{NoSoli}] we use a continuous
driving with the amplitude $A=0.5098$, while the pulse amplitude in
the lower chain can be of the order of $0.004$. As seen such a small
pulse is enough to create a gap soliton and realize amplification
scenario in the oscillator ladder. Returning to dimension units we have $\nu=130.2MHz$ for frequency,
$A=0.73\mu m$ $A^\prime=0.005\mu m$ for driving and pulse amplitude.

Concluding, a clear advantage of the proposed mechanism is that in a
wide range of a weak signal shape and amplitude the generated output
soliton amplitude keeps unchanged providing thus digital
amplification scenario. Moreover, taking into account that we are
using a single operational frequency, the output signal could be
readily used for the further processing. Besides that, different
geometries of the coupled chains could be proposed for implementing
the developed mechanism of amplification for logic gate operations. We considered any number of interacting neighbor units and then applied theory for coupled cantilever arrays. Therefore one has possibilities of studying systems with any precision in terms of number of interacting neighbor units.  

We thank A. Gurchumelia for creating clear visual scheme of a cantilever array [Fig.\ref{Cantilever1}].

\end{document}